\documentclass{svjour3}                     
\smartqed  
\usepackage{graphicx}
\usepackage{slashed}

\begin{document}

\title{Conference Discussion of the Nuclear Force\thanks{Presented at the 21st European Conference on Few-Body Problems in Physics, Salamanca, Spain, 30 August - 3 September 2010}
}


\author{Franz Gross         \and
        Thomas D.\  Cohen \and
        Evgeny Epelbaum \and
        Ruprecht Machleidt
}


\institute{Franz Gross \at
             Thomas Jefferson National Accelerator Facility, Newport News, VA 23606, USA \\
              \email{gross@jlab.org}           
           \and
           Thomas D.\ Cohen \at
              Maryland Center for Fundamental Physics and the Department of Physics, University of Maryland,
College Park, MD 20742-4111, USA \\
\email{cohen@physics.umd.edu}
              \and
           Evgeny Epelbaum \at
              Institut f\"ur Theoretische Physik II, Ruhr-Universit\"at Bochum, D-44780 Bochum, Germany\\
              \email{evgeny.epelbaum@rub.de}
              \and
           Ruprecht Machleidt \at
              Department of Physics, University of Idaho, Moscow, ID 83844, USA \\
              \email{machleid@uidaho.edu}
              }

\date{Received: date / Accepted: date}

\maketitle

\begin{abstract}
Discussion of the nuclear force, lead by a round table consisting of  T. Cohen, E. Epelbaum, R. Machleidt, and F. Gross (chair).  After an invited talk by Machleidt, published elsewhere in these proceedings, brief remarks are made by Epelbaum, Cohen, and Gross, followed by discussion from the floor moderated by the chair.  The chair asked the round table and the participants to focus on the following issues: (i) What does each approach (chiral effective field theory, large $N_c$, and relativistic phenomenology) contribute to our knowledge of the nuclear force?  Do we need them all?  Is any one transcendent?  (ii) How important for applications (few body, nuclear structure, EMC effect, for example) are precise fits to the $NN$ data below 350 MeV?  How precise do these fits have to be?  (iii) Can we learn anything about nonperturbative QCD from these studies of the nuclear force?  The discussion presented here is based on a video recording made at the conference and transcribed afterward.
\end{abstract}

\section{Introduction}  

The discussion session began immediately after Ruprecht Machleidt completed his invited talk.  Franz Gross, the chair, explained that the panel was chosen to include (i) representatives from the four groups who have done precision fits to the low energy $NN$ scatering data, and (ii) experts in chiral effective field theory ($\chi$EFT) and the large $N_c$ (number of colors) limit  of QCD, two ideas that have a strong influence on how we model nuclear forces.  Rob Timmermans had accepted an invitation to represent the Nijmegen group, lead by Johan deSwart, but was unable to come at the last moment.  This left Machleidt and Gross to represent the Argonne, Idaho (formerly Bonn), and Williamsburg-JLab groups.  Evgeny Epelbaum was invited to augment the discussion of  $\chi$EFT and Tom Cohen to discuss large $N_c$. 

The session began with short opening statements from Epelbaum, Cohen, and Gross, followed by discussion initiated by the conference participants.

\section{Effective field theory and the nuclear force (remarks by Epelbaum)} 

One of the most important developments in low-energy nuclear physics in the
past decades has been a general acceptance and popularization of an effective
description of nuclear forces and nuclei motivated by the advent of effective
field theories \cite{Weinberg1990rz}.  The typical nuclear binding energies of the
order of a few MeV per nucleon suggest the description of nuclei in
terms of essentially non-relativistic nucleons. Utilizing the hadronic
picture, the interaction between the nucleons is governed by meson exchange 
and/or nucleon resonance excitations. While it is possible to obtain an accurate
representation of the two-nucleon scattering data below the pion production
threshold in terms of one-boson exchange models of the nuclear force, the
validity of such models is hardly justifiable from QCD. On the other hand, the details of the
short-range part of the nuclear force governed by the exchange of heavy
mesons such as e.g.~$\rho$, $\omega$, $\ldots$ cannot be resolved in
low-energy reactions with external nucleon momenta of the order of a few
hundred MeV/c or less. Eliminating such  redundant information in systems with
a clear scale separation allows for a dramatic simplification of the problem and 
is at the heart of the effective field theory (EFT) approach.   

What are the proper degrees of freedom to address low-energy nuclear
dynamics? The answer depends strongly on the energy range one would like to
describe. The simplest possible EFT emerges from treating 
only nucleons as explicit degrees of freedom. It is justified 
for processes with typical nucleon momenta well below $M_\pi c$ or,
equivalently, for energies well below $ (M_\pi c )^2/m_N \sim 20$ MeV. This is
sufficient to study the properties of atomic nuclei and scattering observables
close to threshold. The resulting approach is commonly referred to as pion-less EFT and
has been successfully applied to explore universality in few-body systems with
large scattering length \cite{Braaten:2004rn}. It
also finds  applications in the physics of cold atoms.

To increase the applicability range it is necessary to
include pions as explicit degrees of freedom. The resulting \emph{chiral} EFT
relies heavily on the approximate spontaneously broken chiral symmetry of
QCD. This symmetry/symmetry-breaking pattern of QCD strongly constrains 
the interaction of pions which play the role of the corresponding Goldstone
bosons. It also implies that pion- and pion-nucleon low-energy observables at  
external momenta $Q \sim M_\pi$ can be computed in a systematic way via
a perturbative expansion in powers of $Q/\Lambda_\chi$,  commonly
referred to as chiral perturbation theory, see \cite{Bernard:2007zu} for a recent review
article. The chiral  symmetry
breaking scale $\Lambda_\chi$ is expected to be of the order of 
$4 \pi F_\pi \sim 1200$ MeV \cite{Manohar:1983md}. In the past two decades,
this approach has been 
extensively applied to the nuclear force problem \cite{Epelbaum:2008ga}. It leads to the picture of
the nuclear force which at large distances is governed by the exchange of one
or multiple pions. In the chiral limit of vanishing quark masses one is
expanding around, these contributions would have an infinitely long range. 
This long-range part of the nuclear force is strongly constrained by the chiral
symmetry of QCD and can be rigorously derived in chiral perturbation theory. 
The chiral expansion for the multiple pion exchange potential is expected to converge
fast at distances of the order of and
larger than the inverse pion mass. The short-range part of the nuclear force
in this picture is driven by physics that cannot be resolved explicitly in
reactions with typical nucleon momenta of the order of $M_\pi
c$. It can be   
mimicked by zero-range contact interactions with an increasing number of
derivatives.  Chiral symmetry of QCD does not provide any constraints for
contact interactions except for their quark mass dependence. 

Considerable progress has been made in recent years in pushing the
calculations of the nuclear forces to higher orders in chiral EFT. 
In particular, nucleon-nucleon scattering has been studied up
to next-to-next-to-next-to-leading order  (N$^3$LO) \cite{Entem:2003ft,Epelbaum:2004fk}, 
see review talk by R.~Machleidt,
yielding an accurate description of the phase shift which is comparable to the
one achieved by modern phenomenological potentials. Systems with three and
more nucleons have so far been analyzed up to next-to-next-to-leading order
(N$^2$LO) \cite{Epelbaum:2008ga}. 
The results for most of the low-energy nucleon-deuteron scattering
observables are in a good agreement with the available experimental data 
with the exception of the vector analyzing power in elastic scattering, the so-called 
$A_y$-puzzle, and the cross section in the symmetric space star configuration
in the breakup reaction \cite{Epelbaum:2008ga}. Given an extremely strong sensitivity of $Nd$ $A_y$ 
to $NN$ triplet P-wave phase shifts, one expects a rather large
theoretical uncertainty for this observable. Thus, 
the $A_y$-puzzle at N$^2$LO appears less
worrisome than the disagreements in the breakup. The N$^3$LO
corrections to the three-nucleon force are currently being worked out and
implemented numerically in the scattering codes
\cite{Ishikawa:2007zz,Bernard:2007sp}. 
Electromagnetic currents are
also being studied in the framework of chiral EFT by the JLab-Pisa \cite{Pastore:2008ui}
and Bochum-Bonn-J\"ulich \cite{Kolling:2009iq}
groups. These studies provide an extension of the
pioneering calculation by Park et al.~\cite{Park:1995pn} restricted to very low
photon momenta and will allow to test the theory in low-energy photon-induces
reactions. 

An important conceptual issue that is being debated by the community is related
to (the meaning of) the non-perturbative renormalization of the Schr\"odinger
equation in the context of chiral EFT. While
renormalization is carried out straightforwardly in chiral perturbation theory
by absorbing all ultraviolet divergences that appear at a given order
into a redefinition of (a finite set of) low-energy constants, things are
more subtle in the few-nucleon sector. In particular, an infinite number of
nucleon-nucleon counter terms are needed to absorb ultraviolet divergences
that emerge from iterating the static one-pion exchange potential in the
Lippmann-Schwinger equation. Non-availability of analytical results for the 
scattering amplitude provides another complication. A rather plausible and 
presently most frequently used approach was suggested by Lepage \cite{Lepage:1997}
and relies on employing
a finite cutoff to regularize the Lippmann Schwinger equation. 
Renormalization is carried out implicitly  
by adjusting the two-nucleon contact interactions to fit low-energy data for each
given value of the cutoff. Self-consistency can be verified a
posteriori employing the so-called Lepage plots. The role and the choice of
the cutoff parameter as well as the most efficient and consistent way to organize
the EFT expansion for nucleon-nucleon scattering are still under
debate. A closely related question 
concerns the identification of the breakdown scale of the chiral expansion for the
nuclear force. A careful look at the two-nucleon potential emerging from  
pion-nucleon rescattering diagrams reveals that (i) pion loop contributions are
enhanced by factors of $\pi$ indicating that the usual estimation for the
breakdown scale $\Lambda_\chi = 4 \pi F_\pi \sim 1200$ MeV is too
optimistic in the nucleon-nucleon sector and (ii) the partially resummed multi-pion exchange potential
features poles at distances of the order $0.5 \ldots 0.8$ fm. The appearance
of these poles indicates that the chiral expansion of the pion-exchange
potential cannot be trusted at these or shorter distances anymore and provides
an estimation of the breakdown distance scale.  

This raises an important issue on the possibility of improving
the convergence of the chiral expansion for the long-range part of the nuclear
force. The developments outlined above are based on the effective Lagrangian
which involves pions and nucleons as the only explicit degrees of freedom. On
the other hand, the $\Delta$(1232) isobar is well known to play an important
role in nuclear physics due to its low excitation energy and strong coupling
to the $\pi N$ system. In the present formulation, all effects of the $\Delta$ 
isobar are taken into account implicitly through the values of certain 
low-energy constants in the effective Lagrangian. The \emph{explicit} inclusion of
the $\Delta$ in the EFT by treating the $\Delta N$ mass splitting as an
additional soft scale \cite{Hemmert:1997ye} allows one to resum a certain class of important
contributions leading to an improved convergence. It, however, also requires 
considerably more involved calculations. 
The first contributions
of the $\Delta$ to the two-nucleon force appear at NLO and were worked out by
Ordonez et al.~\cite{Ordonez:1995rz}
and Kaiser~et al.~\cite{Kaiser:1998wa} a long time ago. Recently, Krebs, Mei{\ss}ner
and myself derived  the subleading contributions at N$^2$LO and 
confirmed an improved convergence of the EFT expansion
\cite{Krebs:2007rh}. 
The N$^3$LO contributions of
the $\Delta$ isobar are not yet available. At this order one expects, in particular, large
corrections to the three-nucleon force whose description in the  $\Delta$-less
formulation would require to go  beyond N$^3$LO.

\section{The nuclear force and the large $N_c$ limit of QCD (remarks by Cohen)}
\label{intro}

The  $1/N_c$ expansion large $N_c$ limit of QCD \cite{tHooft:1973jz,Witten1979kh} has proven to be a very valuable tool in hadronic physics.  The idea was originally introduced to provide a useful expansion to describe hadronic physics  at low momentum where the perturbative expansion fails. The key idea is that QCD  in a many-colored world is qualitatively similar to a world in which the number of colors, $N_c$ is three.  If this is the case, then it is sensible to use the the large $N_c$ limit as a starting point and then use $1/N_c$ as the basis for an expansion.  At a field theoretical level it is easy to understand some of the key features of the large $N_c$ limit.  In particular, it is simple to see that planar diagrams dominate and that correlation functions with fixed quantum numbers are dominated by diagrams with the fewest number of quark loops.  The study of mesons is particularly straightforward as it is based directly on the study of the correlation functions.  Baryons   are much more complicated in that the number of quarks in a baryon itself grows with $N_c$ potentially leading to  complicated combinatoric factors.  Witten\cite{Witten1979kh} showed, however, that baryons can be understood at large $N_c$ from the perspective of mean-field theory.

The $1/N_c$ expansion has proven to be a very useful tool in hadronic physics.  It leads to a qualitative and, in some cases, a semi-quantitative understanding of many hadronic phenomena.  For example, the OZI rule becomes exact at large $N_c$.  Of particular importance for the present purpose is the fact that at large $N_c$ baryons have an emergent contracted SU(2$N_f$) symmetry\cite{GS84a,GS84b,DM93a,DM93b,DJM94,DJM95}. As result of this emergent symmetry point  at large $N_c$ baryons fall into  multiplets of  degenerate baryons with spin equal to isospin.  Members of these multiplets (such as the nucleon and delta) are split due to $1/N_c$ corrections.  Moreover, up to $1/N_c$ matrix elements of operators between baryons are given by Clebsch-Gordan coefficients of this group times reduced matrix elements which are universal in the sense that they are the same for all matrix elements in the multiplet.  Thus large $N_c$ makes concrete predictions for baryons.  Generally it works rather well with $1/N_c$ corrections of the characteristic size one expects.

Of course, the usefulness of such an expansion depends on  how rapidly it converges (or in the event it is asymptotic, how rapidly does it start to accurately reflect the correct value).  One does not expect the expansion to give much predictive power unless the coefficients characterizing the expansion are  ``natural''; otherwise, it is hard to argue that the neglected terms associated with truncations are small.  It is not obvious that the quality of the expansion is necessarily the same for all observables.  It is plausible that coefficients for some observables might be natural (and the expansion useful) while for other observables this is not the case.  The success of the $1/N_c$ expansion in describing many hadronic observables suggests that the expansion often has natural coefficients when applied to hadronic physics.  Does this suggest that one can simply apply these methods to nucleon-nucleon interactions?

{\it A priori} the answer is ``no''.  Characteristically the energy scales of relevance in nuclear physics is much smaller than in hadronic physics---and for reasons having nothing to do with large $N_c$.  For example, the binding energy of the deuteron is order $N_c^1$ (using the standard analysis of the type pioneered  by Witten) but it is only 2 MeV.  In contrast the mass  difference of the nucleon and the delta which is order $1/N_c$---two orders down in the expansion---is nearly 150 times larger.  This strongly suggests that in nuclear physics the coefficients are not natural.  This in turn suggests that a direct application of the $1/N_c$ to phenomena at the nuclear scale might be useful in a world with $N_c$ in the thousands, it is likely to fail in our world where $N_c=3$.

This need not mean that large $N_c$ analysis is useless for the world of $N_c=3$.  Indeed, in an important way the problem is reminiscent of chiral perturbation theory.  Recall that chiral perturbation is based on a scale separation between the pion mass (or external momenta) and the natural hadronic scale of $\sim$1 GeV.  Chiral perturbation theory has proven useful in hadronic physics.  However, it is clearly inappropriate in  direct calculations of nucleon-nucleon scattering amplitude.  The fact that scattering lengths (which do not diverge in the chiral limit) are much larger than the inverse pion mass means that chiral perturbation theory has clearly broken down for direct low energy nucleon-nucleon interactions.  Nevertheless there has been a significant amount of work in trying to apply chiral perturbation theory to nucleon-nucleon forces.  There have been many variants on how to do  this, but one important approach is that pioneered long ago by Weinberg (for a review see \cite{Weinberg}) in which systematic chiral power counting is applied to  the nucleon-nucleon potential which is then used to compute amplitudes.  To the extent that this reasoning is legitimate, it can equally be applied to large $N_c$ analysis.  That is, one may argue that even if one is in a regime in which the  $1/N_c$ expansion has broken down for the two nucleon observables, it may still be appropriate for nucleon-nucleon potentials.  Of course, in fairness one must add at this point that it is something of an {\it ad hoc} assumption that this is the case for either  the chiral expansion or the $1/N_c$ expansion.

If one accepts that the expansion is useful for the potential, then one can use the $1/N_c$ expansion and the large $N_c$ limit to get at least some insight into nucleon-nucleon interactions.  The key point is that contracted SU(2$N_f$) symmetry applies to the nucleon-nucleon interaction as well as nucleons\cite{KM}.  This constrains which terms contribute to the nucleon-nucleon potential at leading order ($N_c^1$); these include  the isoscalar central force or  the isovector tensor force; other components of the interaction turn out to be down by powers of $N_c^{-2}$ compared to the leading one.  Now, the nucleon-nucleon potential is not directly an observable; different potentials lead to the same observables thus one might worry as to whether qualitative predictions for the characteristic strengths of the different components of the potential  obtained from the emergent symmetry are meaningful.  Nevertheless, if one simply compares the qualitative predictions of the emergent symmetry obtained at large $N_c$ with ``realistic'' potentials fit to nucleon-nucleon phase shifts, one finds good qualitative agreement: the terms which should be large are large and the terms which should be small are small.

The large $N_c$ limit is also useful in another way. One might worry that a meson-exchange picture for nucleon-nucleon forces is hard to understand from the perspective of QCD.  Large $N_c$ provides a tool that gives insight into the consistency of meson exchange models.   In particular, the patterns seen in the potential due to the contracted  SU(2$N_f$) symmetry emerge simply in a one-meson exchange model, with the meson-baryon couplings given from large $N_c$.  Moreover, multi-meson exchanges turn out to be inconsistent with the underlying large $N_c$ counting rules unless remarkable cancelations occur.  However the emergent symmetry enforces precisely the needed cancelations\cite{BCG,BC02}.

\section{Relativistic theory and the nuclear force (remarks by Gross)}

Relativity is an exact symmetry of nature and should be part of any description of the nuclear force.  In the context of $\chi$EFT it is often argued that relativistic effects, expected to be of the order of $(p/M)$, are incorporated order by order in the perturbations series.  Alternatively, it is possible to use one of the many methods that are explicitly covariant.   Could a manifestly covariant description give us a new perspective on this old problem?  

Some problems {\it require\/} relativistic theory.   For example, high $Q^2$ elastic electron-deuteron scattering studies the transition from a deuteron at rest to one recoiling with high velocity.  Perhaps much of the physics  can be explained by simply doing the relativistic boost of the final state correctly (to {\it all\/} orders).  But even when there are no large momenta a covariant description might enjoy some advantages.  I am reminded of lessons learned from the Dirac equation.  Before its discovery, atomic physicists treated many effects independently: the $-p^4/(8m^3)$ term from the relativistic mass increase, the Darwin term $e\,\nabla^2 \phi/(8m^2)$, the spin-orbit term $e\,(d\phi/dr)\, \sigma\cdot{\bf L}/(4m^2r)$, and the Zeeman effect $-e\,{\bf B}\cdot({\bf L}+2{\bf S})$ (where the Dirac equation {\it predicted\/} the mysterious factor of 2 that multiplies the spin, ${\bf S}$).  Maybe a treatment that retains the full Dirac structure of the nucleons would automatically include many small contributions to the $NN$ force that would be difficult to know about and to insert by hand?  If this were true, retaining the full Dirac structure would give an efficient description of the data, with fewer parameters.  This is precisely what we find.

The equations of the Covariant Spectator Theory (CST) are manifestly covariant and conserve four-momentum in all intermediate states \cite{Gro69,Gro74,Gro82,Gro82b,Gro92,Sta97b,Sta97}.  In the two and three-body sectors, the CST propagator restricts all particles but one to their positive energy mass-shells, retaining the full Dirac structure of the remaining off-shell particle.  The generalized Pauli principle is preserved by explicitly antisymmetrizing  the kernel.  As currently applied, the relativistic kernel is approximated by a one boson exchange (OBE) model \cite{Gro08}.   We found two models that gave precision fits to the  2007 $np$ data set (3788 data) below 350 MeV.  One model, WJC-1 with 27 parameters has a $\chi^2/N_{\rm data}$ of 1.06. Here we allowed the masses of the heavy bosons and most of the coupling constants to vary to obtain the best fit possible, and the fit is comparable to the best fits ever achieved. A second model, WJC-2, has only 15 parameters and was simplified as much as possible by fixing some of the meson masses and eliminating some of the less important degrees of freedom.  This model has a $\chi^2/N_{\rm data}\simeq$ 1.12, remarkable for such a simple model.  Both of these are true OBE models, with the same OBE parameters for all partial waves.  These models have fewer parameters than previous precision fits to the data, and WJC-2, with only 15 parameters, is more efficient that the $\chi$EFT models which require 24 parameters when calculated to N$^3$LO.  Perhaps this efficiency comes from retaining the full Dirac structure of the off-shell particle.

Use of the CST to describe two-body scattering  in OBE approximation was originally justified by observed cancellations between ladder and crossed ladder diagrams that occur when one of the two particles is  neutral and massive (leading to the one body limit) \cite{Gro69,Gro82}.   It now appears that the large $N_c$ limit of QCD leads to similar cancellations in the realistic case of nucleons and $\Delta$s exchanging mesons, and it is therefore quite possible that OBE has a deeper justification coming directly from QCD.

In any case, the CST not only provides an efficient and accurate description of $np$ scattering, but it also provides some remarkable insights.  It was observed a long time ago \cite{Gro74} that the CST can provide a plausible explanation for the repulsive core.  Decomposing the off-shell nucleon into positive and negative energy contributions  leads to a set of equations that couple these two channels.   In the nonrelativistic limit, the equations become
\begin{eqnarray}
\left[\frac{\nabla^2}{m}+E\right]\psi^+&&=V^{++}\psi^++V^{+-}\Psi^-
\nonumber\\
\qquad 2m\,\psi^-&&=V^{-+}\psi^++V^{--}\psi^-\, .
\end{eqnarray}
Eliminating the negative energy wave function, $\psi^-$, gives a Schr\"odinger equation for $\psi^+$ with an effective potential
\begin{equation}
V_{\rm eff}=V^{++}+\frac{|V^{+-}|^2}{2m-V^{--}}\,.
\end{equation}
The second term is always repulsive and of the general size and shape required to reproduce the repulsive terms in phenomenological potentials; it complements the repulsion due to $\omega$ exchange leading to a smaller $\omega NN$ coupling constant closer to estimates from SU(3).

\begin{figure}
  \includegraphics[width=2.5in]{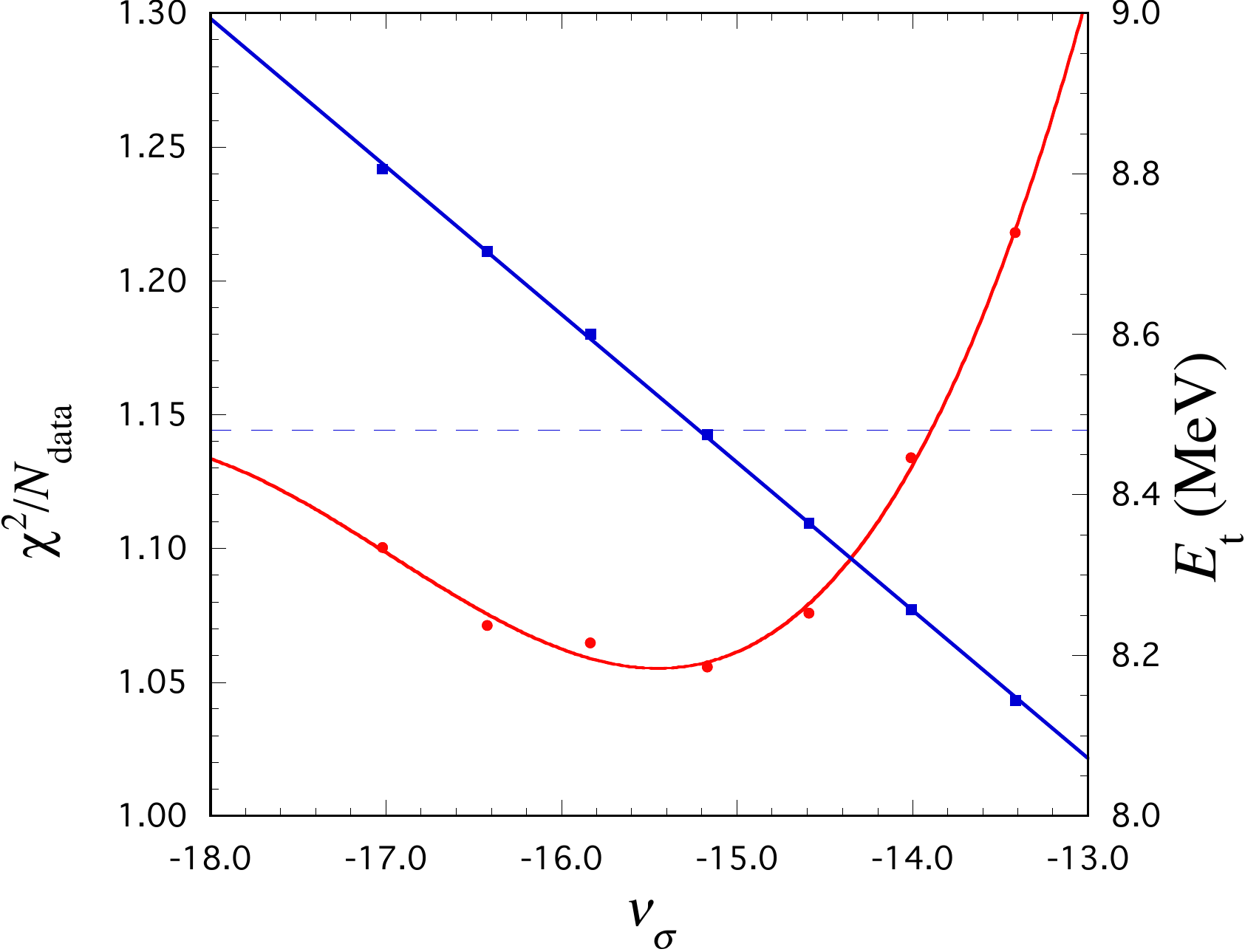}
\hspace{2pc}
\begin{minipage}[b]{12pc}\caption{\label{fig:1} The family of WJC-1 models with $\nu_\sigma$ constrained to various fixed values. The left-hand axis shows the best $\chi^2/N_{\rm data}$ that can be found for each value of $\nu_\sigma$ (the data shows some
scatter with respect to the solid line, which is a cubic fit to the 7 cases
shown) and the right-hand axis shows the triton binding energy ($E_t$ in MeV) for each member of the family. Note that the correct binding
energy (shown by the dashed horizontal line) is obtained for the value of $\nu_\sigma$ that also gives the best fit to the data.
\vspace{1pc }
}
\end{minipage}      
\end{figure}

Furthermore, recent CST models all predict the {\it correct\/} binding energy for the triton {\it without any three body forces\/} \cite{Sta97,Gro08}.  This requires that the covariant $\sigma NN$ coupling for incoming (outgoing) nucleons with four-momentum $k$ ($p$) include an off-shell term
\begin{equation}
\Gamma_{\sigma NN}(p,k)=g_\sigma {\bf 1}-\nu_\sigma \left[\frac{m-\slashed{p}}{2m}+\frac{m-\slashed{k}}{2m}\right]
\end{equation}
with the off-shell parameter $\nu_\sigma$ allowed to vary during the fits to the two-body data.  The value of $\nu_\sigma$ that gives the {\it best fit\/} to the data also gives the {\it correct\/} binding energy, as shown in Fig.\ \ref{fig:1}.  This remarkable behavior is not an accident; it occurs also for the family of WJC-2 models and for a family of older models used in the original calculation \cite{Sta97}.

To appreciate the significance of this result I emphasize two points.  In the CST the OBE mechanism, when extended to three nucleons, generates {\it no\/} three body forces (because the summation of {\it all\/} OBE diagrams leads exactly to the Faddeev sum of two-body scattering amplitudes, with {\it no\/} additional three-body force diagrams).  Secondly, contributions from the off-shell couplings will cancel the nucleon propagator shrinking successive interactions to a point.  To obtain the same results from a model {\it without\/} off-shell couplings, we would have to add an infinite number of multiple boson exchange loop diagrams, and in the three-body sector also an infinite number of three-body force diagrams (some involving loops).   Looked at from this point of view, off-shell couplings are an efficient way of including infinite numbers of  complicated two and three-body force diagrams, with their couplings all fixed by the small number (four) of off-shell parameters in the model.  
The remarkable fact is that the three-body force diagrams generated in this way give exactly the correct three-body binding energy.  

These ideas are discussed further in the literature and in the conference talk by Alfred Stadler.  Our conclusion, in the context of this discussion, is that OBE models in CST are important alternatives to nonrelativistic phenomenologies and to models based on $\chi$EFT.

\section{Discussion}

The chair opened the discussion by  reminding the audience of the three questions posed for the conference (and restated in the abstract of this report), but asked the members not to be constrained by these questions, but to address issues in a manner most comfortable for them.  Participants were told that the discussion was being video taped  and were asked to speak clearly into the microphone.

{\bf Alejandro Kievsky} (INFN Pisa): Listening to the talks I have the impression that the speakers believe that once the transition has been made from QCD to nuclear physics, the rest will come naturally.   This is a new point of view.  Many years ago I remember another round table at Evora.  There we assumed that we had a potential, and few-body physics would help us understand if the potentials were O.K. 

Now it seems that few body physics is trying to understand the transition from more fundamental theories and then assumes that nuclear physics will come naturally.   Well I think it is not completely correct in all aspects. It seems too optimistic -- nuclear physics is very difficult and we need potentials of course, and we need precision, but this is not enough; all ingredients are not in sight. For example, I saw in the presentations that we look too much to agreement with two nucleon phase shifts, but nothing is said about the agreement with the phase shifts in the three-nucleon systems or in the four-nucleon systems. Why are we missing this part?  Why don't we want to fit the three nucleon phase shifts, for example? This is, in my opinion, a new step that we need to look at.  It seems that if we do well fitting the two nucleon system with chiral perturbation theory, or with relativistic theory, it is not necessarily true that we will fit the three nucleon systems. So we need to include more ingredients into the methods. I don't know if I was clear.


{\bf Peter Sauer} (Leibniz University Hannover): Presumably Alejandro used different words for what I want to say also. I must admit that when I'm looking at your first question, I'm backing off.  I'm interested in nuclei and my interest in nuclei always was that there must be microscopic building blocks that describe the interaction of two nucleons, of three nucleons, and from them I want to derive the properties of nuclei.  O.K., you may say, Peter Sauer, you are from the stone age of few body physics. That's not our interest anymore. It was my interest and perhaps now you have different ideas.  But if I'm coming back to my original idea -- why I love few body physics -- it is an intermediate step in going to the understanding of heavier nuclei.   Then, of course, I want to understand potentials and  I have to do quantum mechanics in order to go from two, to three, to four, to five nucleon systems. And therefore when you talk about relativistic phenomenology I would be all with you but presumably I would vote for relativistic quantum mechanics. I do not know if any of the other items which you have there would help me in my understanding of nuclei. So that is my worry about the first question.

{\bf Gross}: Well, this first question is not meant to be part of my presentation. This is meant to be for the whole group to discuss. I should, perhaps, answer you briefly. You are absolutely right. The stuff I do is very limited in its applicability. When I started I didn't know how to do three-body systems. Now we can do that. Alfred has figured out how to do that. We probably can do more complicated systems.  But don't direct all the comments to me just because I was the last speaker; most of the work has been directed toward understanding complicated nuclei and much progress has been made.

{\bf  Enrique Ru\' iz-Arriola} (University of Granada):    I would like to address one of the questions. You know, there is a big difference between explaining an experiment and understanding the theory. As a theoretician I want to understand theory with the guidance of experiment; of course we want to describe experiment. Regarding nucleon-nucleon scattering, we have had a marvelous phase shift analysis to go on for the last fifty years. Now some of us want to see what connection there is between these kind of fits and anything resembling QCD.   And of course, during the years people have been trying to see the smoking gun signature of quarks, but actually, as Tom Cohen has  properly pointed out, the large $N_c$ limit  gives  us  a model independent  way of testing the quark model, in a sense.  

Also, people are talking about potentials. In the large $N_c$ limit the well defined thing is the potential. That is not the case in chiral perturbation theory. In chiral perturbation theory the potential does not appear naturally. But in the large $N_c$ limit the potential has a well defined meaning in the sense of a generalized Born Oppenheimer (still it is not exactly that).  So what are the requirements.  Of course, we'd like to fulfill chiral perturbation theory, large $N_c$,  relativity, phenomenology. Now, for this kind of problem, if you have to be confronted, suddenly, with the need to reproduce the two body, three body, four body, and all kind of reactions, I think it is hopeless. So some of the issues can be addressed in a simplified system; for example, in the nucleon-nucleon system, where we can study large $N_c$.

How do we identify a symmetry?  How do I prove or disprove the statement  that chiral perturbation theory works in the nucleon-nucleon interaction?  The potential already gives you an acceptable $\chi^2$ so how do you validate or invalidate it? This is one of the issues. 

I would like to say a few words about another issue addressed in one of the talks: whether or not we are removing the cutoff. Of course, in a mathematical sense, maybe it is science fiction to try to remove the cutoff. One of the reasons people renormalize is that they identify the infinities in this way. Once you have a handle on your infinities then you can expect a smooth dependence on the cutoff.  And we were shown some results  where you have convergence, but not to experiment. That is not bad for the renormalization. I think it is bad for the N$^3$L0 potential.  Some might say that this is bad for the renormalization. I say no. My conclusion is that the potential has only two pion exchange and that the physics has to include three pion, four pion exchange and all that. So it would be nice to agree on the conclusion because the result would be the same. It will not change. So from this point of view the issue is very tough but confronting this question  is not just a way of wasting time.  The reason to make a good theory is to make a prediction when you cannot make any experiment beforehand. 

{\bf Gross}: Tom, will you comment on that?.

{\bf Cohen}: Well, I just wanted to make a brief response. I think large $N_c$ does have some value in connecting some aspects of QCD to data. However, you made another point which was that you ought to be able to know what those aspects are a priori. And in point of fact, with large $N_c$ that's not always the case.   I mean, in large $N_c$ the $\eta'$ should be a Goldstone boson. It isn't close to it. You say, oh, that's just an effect of anomalies and so forth, topological susceptibilities. But the point is, it's not a completely compelling story because after all,  $N_c$ is only three. 

Regarding your other comments about cutoffs, I sort of agree with you that if you have a theory which is truly a theory and the whole idea is that you are completely insensitive to the details of short distance physics, then it should be the case that you can take the cutoff to infinity in a completely innocuous way because you're insensitive to what's going on up there if the expansion is converging properly. And if you can't, and it was working very well with a small cutoff of a GeV or so, and fails with an infinite cutoff, it somehow suggests that what's happening at very high scales is, in fact, effecting your observables and that somehow this notion of complete scale separation underlying effective field theory may be in some trouble.

{\bf Giuseppina Orlandini} (University of Trento): The way I have always seen few body physics is as a bridge between QCD and nuclei.  So in that respect, I agree with Sauer's point of view that, O.K., we make an effort to build this bridge.  That means give me a potential, I develop technologies to calculate observables in many-body systems and few-body systems and if I am successful with this kind of potential that somebody gives me then I say, O.K., everything is fine. I can proceed probably to larger systems and maybe I can predict many-body behavior, collective effects, mean field effects and so on.  Answering that question from this point of view is hard. Few body physics is a bridge and what do each of our modern ideas contribute to our knowledge of the nuclear force?   I would re-phrase that -- what do our modern ideas contribute to our knowledge of the nuclear force as used in nuclear physics? 

Certainly all those ideas contribute because what we want to learn is how to use QCD.  I  mean, we cannot do an ab initio  calculation using QCD, but  there must be some relevant effective degrees of freedom,  the real effective degrees of freedom relevant for nuclear physics, and we want to know what these are. So chiral perturbation theory is fine, yes, and nucleons and pions are good; O.K., let's accept it.  And maybe large $N_c$ gives effective degrees of freedom which rule everything.   Or, maybe it's relativity with pions and nucleons which works.  Then few-body physics can answer which of these three ideas (or others) are more relevant for nuclear physics.   But the perspective, I agree, is still towards the many-body system.

{\bf Jean-Marc Richard} (University of Lyon): I have a question for the first speaker.  You did not mention any of the earlier work on the nucleon-nucleon interaction, some that you did yourself and was a very great contribution.  At that time the potential was based, roughly speaking, on boson exchange between nucleons. Now that you have developed a more successful potential, can you look back into the past and tell us what was missing in the early potentials?   Maybe not enough pion exchange, too large a  $\rho$ coupling, missing high-spin meson exchange?   It is interesting to see how this has been improved.   Maybe this is a lot of work; to project out the $t$-channel, see where the cuts and the poles are, and to make the link between the past and the present work.

{\bf Machleidt}: A very good question. In practice, the meson theoretic potentials like the famous Paris and Bonn potentials, when you compare them with the chiral potentials,  contain, in  fact and phenomenology, essentially the same physics. There is a good reason for that.  When two different types of potentials, no matter what their background,  describe $NN$ scattering, they have to describe the same facts of $NN$ scattering and the $NN$ scattering data determine how strong the spin-orbit  force is, how strong the tensor force has to be, how strong the central force has to be, and the data are so good there isn't much latitude. So if you approach this discussion from a phenomenological point of view,  the contents of the old meson theoretic potentials and the new chiral potentials are essentially the same except they extract it from different starting places.   If you want to do quantitative, accurate, few-body physics, and are not interested in the fundamentals, you could say it doesn't really matter which one you take.  As a matter of fact, the practitioners in few-nucleon or many-body physics presently use a  chiral potential 50\% of the time and something like CD-Bonn or  Nijmegen potentials  50\% of the time and the results are not that different.  So to summarize in a short and concise way, the basic properties of a quantitative nuclear force have to be, in both cases, the same, but the theoretical origin is slightly different. 

There is another bridge.  It has been shown how certain contributions to the chiral theory are really equivalent to certain meson exchange contributions (this has been called resonance saturation).  So there is (I wouldn't say  a one-to-one correspondence, that's a little bit too much) a very intimate correspondence between the two, and it has to be that way because two   accurate and quantitative potentials cannot be too different from each other.

{\bf Cohen}: But, Ruprecht, if you believe that then it raises the question -- O.K., so in what way do you actually make any money by developing this marvelous chiral potential if all you are doing is fitting the phase shift data, why don't you just take a bunch of Gaussians and be done with it.

{\bf Machleidt}:  Of course there are many-body forces.  Chiral theory is better because it generates many body forces in a systematic way, and conventional theory does not allow you to do that in that systematic way.

{\bf Epelbaum}: Actually, I would also like to add a couple of arguments. As long as you are only interested is fitting nucleon-nucleon data, then basically it doesn't really matter what interaction you take as long as it reproduces data with a $\chi^2/N\sim 1$.   The real advantage, or promise, of chiral perturbation theory comes into play if you want to simultaneously describe low energy dynamics including, let's say, pion reactions like pion scattering on light nuclei, or if you switch on external sources and consider processes with photons or neutrinos. The chiral approach gives you immediately  strict rules for how to compute all the ingredients you need in the calculations -- currents and many body forces.  And it is a very convenient tool if you want to study symmetries; for example, isospin symmetry, charge symmetry breaking and so on.  You start immediately from the beginning with a Lagrangian which respects the corresponding symmetries; it's just an ideal tool. And of course, if you want to ask questions about the quark mass   dependence, I believe that this framework is also extremely useful. This will be more and more important with the lattice data that are coming.

{\bf Roman Kezerashvili} (CUNY): I am, somehow in the same boat as some speakers here and I would like to say everything depends on what your starting point is.  What is your initial hypothesis? If you describe nuclear systems with classical quantum mechanics you have  your Schr\"odinger equation, God gave you some potential, you plug this potential into the equation, you solve the equation and your conclusion must be within your hypothesis. Never overestimate your results and give a conclusion that is beyond the initial assumptions. If you are building a relativistic theory, you have to describe atomic nuclei also within the theory and make the conclusion which comes within the theory. In relativistic approach you have no two, three, four body problem, you have zero body problem basically, and you don't need a potential at all. But there is a third approach, the quasi-potential approach which was developed some time in the 70s-80s so here you have the same  Schr\"odinger equation with quasi-potential or Schr\"odinger-like  Bethe-Salpeter equation with quasi-potentials, or Bethe-Faddeev equations with quasi-potentials, and you can make conclusions there also within these quasi-potentials. 

{\bf Yury Uzikov}  (JINR, Dubna):  I would like to note here first that, of course, chiral symmetry of classical QCD Lagrangian is very important for hadronic systems and for nuclei, and it is especially important that it is broken on the quantum level. The quantum level has a chiral condensate which determines hadron masses and much of the general properties of hadrons and nuclei. There is a statement by Dmitri Diakonov made twenty years ago that nuclear physics would be different if chiral symmetry were not broken or if it were absent. Of course it is very important direction to derive nucleon-nucleon forces from this broken symmetry. But, it seems that the language which is used, field theory, is not quite adequate for the problem because field theory assumes point-like objects, while real nucleons and mesons have finite size and therefore we have a problem, seen today and yesterday with normalization.    But  I do not see how to avoid this. The problem with nucleon-nucleon force is not solved yet and will not be solved in chiral perturbative theory because the scale of broken symmetries is one GeV but a core is visible in the transverse momentum of 0.8 GeV.   It will not be possible to understand core within in this approach.

{\bf Epelbaum}: O.K., so just quick remarks.  Nucleons are point like if you will look at them at extremely low energy so it is basically a matter of scale.  Of course, if you increase the energy then you are leaving the domain of validity of  this approach.

{\bf Machleidt}: To put it shortly, the answer to your question is you are mixing two issues. In chiral perturbation, nucleons and pions are point like. Period. 

{\bf Uzikov}: Clearly they are not point like ...

{\bf Machleidt}: It must come in some higher order correction or so but that's the concept. If you find it realistic or not ...  then O.K.  Point-like particles have to be part of the concept of chiral perturbation theory, otherwise the whole thing has no basis.

{\bf Cohen}: That is the statement that chiral perturbation theory is an effective theory Ð it's not a field theory. It's a field theory that is based on the scale separation which is only good for describing low energy observables, and to the extent that pions are considered much lighter than nucleons and other degrees of freedom like $\rho$'s, then it is a systematic expansion but its only an expansion. It is not a fundamental theory and so the issues involving finite size are described at higher orders.  So I don't think there is a conceptual question. The only question is a practical one. Is the scale separation big enough to be useful, etc., etc. But from an intellectual point of view the whole notion of effective field theory lets you avoid exactly that question.

{\bf M.\ Teresa Pe\~na} (Instituto Superior Tecnico, Lisboa): What I want to do now is make a few comments  in answer to the first question.  These are based on my personal feelings and my personal evolution over a long twenty year period, witnessing development after development on our knowledge of $NN$ and $3N$ force. Let me start by saying that what we are facing here is the problem of having two different audiences and two different types of interests. Some of us want  to go to the deep origins and the underlying QCD problems related to the $NN$ interaction. Others are more pragmatic and want  to describe some nuclear reactions and some nuclear observables. In a sense the objectives of those two audiences are different. But let me stress some aspects that were important in my experience. 

I started doing pion production reactions and problems related to the pion coupled to the $NN$ channel. I must say that from the very beginning I was very much aware that in the pion production problem the two scales -- the soft scale and the hard scale -- couple, and  I thought, to myself, that this was the main problem of nuclear physics -- that these two scales could not be disentangled and treated  separately in a clear way. So I was very happy when chiral perturbation theory first appeared because I thought there was hope to set a cutoff and really separate the two energy ranges and to see clearly where one starts and the other one ends. Unfortunately, as clearly illustrated in the Machleidt's talk this morning,  there are still unsolved problems in that respect. 

But there has been huge progress with the development of chiral effective theory.    I still remember years ago how pleasant it was to discover that the Tucson-Melbourne three-body force, the very old three-body force that people where applying to nuclear calculations, had a problem that was pin-pointed precisely thanks to the development of chiral perturbation theory.   That was gratifying and, for me, intellectually an advance. 

With respect to the four approaches that are the focus of this panel, I must say that I have the feeling that they are able to answer different problems. This is my opinion. I like the field theoretical relativistic approach because in my view it can be extended to calculations of the electromagnetic current in a manner consistent with the $NN$ interaction. With respect to the large $N_c$ limit, its appeal is, of course, the connection to QCD and the possibility that it may connect the $NN$ force to other symmetries beyond chiral symmetry that are present in QCD. This is what I wanted to say from my experience.

{\bf  Werner Tornow} (Duke University and  TUNL):  As an experimentalist, I am of course very much concerned about the fact that the community does not know much about the neutron-neutron interaction. I mean they know the interaction very well at low relative energy. I believe that the neutron-neutron scattering is, in fact, $-$18.64 fm (or something like this).   But I really question whether we know, accurately, the neutron-neutron interaction in the energy range of, say, 10 - 20  MeV.   We have all done phase shift analysis for proton-proton scattering and neutron-proton scattering, and we learned a lot from this but, of course, we have made errors as well. For example, you may remember that when the  Nijmegen  group came up with their phase shift analysis for proton-proton scattering and neutron-proton scattering, they forgot about the magnetic moment interaction and we had to tell them they had better include it; that it has a big effect.  And when we did the first three-body  phase shift  analysis  proton-deuteron scattering, neutron-deuteron scattering, we also did not include the magnetic moment interaction. We all know now that it plays a very important role and that from the very beginning we should do things right. But the point I'd like to make now is what we can do to learn a little bit more about the neutron-neutron interaction.  We need to do phase shift analysis of three-body systems, breakup systems, not elastic scattering. We don't learn much from elastic scattering but we should do them for the breakup channels. That's what Peter Sauer pointed out a long time ago. 

More importantly, it is about time now to do the phase shift analysis for the four-body system. Again we do not have a complete set of data yet --  we have $p\,^3$He, $p\,^3$H,  and $n\,^3$He; $n\,^3$H is missing.  Nevertheless you will see, and we all know, that if you look at the analyzing power, for example, of these four body systems, the pure isospin systems like $p\,^3$He very much disagree with theory unless you now add the three body force in N$^3$LO.   From the very beginning the $n\,^3$He analyzing  powers don't need the three body force -- from the very beginning they already come very close in comparison to theory.  I think these questions are all related and we have to learn a little bit more how to extract information from these systems about the neutron-neutron interaction at finite energy. 

{\bf  Machleidt}:  I believe that chiral perturbation theory is a very accurate and very systematic tool for pinning down all forms of charge symmetry breaking.  Dr. Epelbaum is a much greater expert on this and he would probably confirm that. It is true that, historically, big mistakes were made.   But even if the Nijmegen group made a mistake twenty years ago it doesn't mean that this will happen again and again. So the bottom line is: I'm confident that we have a pretty good knowledge of charge symmetry breaking, and so starting from the $pp$ interaction which we know very precisely I think we can, with great confidence, draw accurate conclusions about the $nn$ interaction.

{\bf Jurij Darewych} (York University, Toronto): I would like to make a couple of remarks relevant first to question 1: the use of relativistic equations or phenomenology. I think that is important. I'm reminded, for example, of very low energy polarized electron or positron scattering in atomic physics where the use of the Dirac equations which seems completely unnecessary at such low energies is, in fact, very advantageous for explaining many phenomenon. There are caviats, of course. Relativistic calculations would ideally start with Bethe-Salpeter equations which are notoriously difficult to solve beyond a couple of particles.  In practice, one needs to do single-time reductions. Franz' equation is one such, but these are non-unique and that must be kept in mind. 

The second point I'd like to make refers to the connection between QCD and the nuclear potential. If we take the analogy that the internucleon potential is kind of a van der Waals force then again, with the atomic physics analogy, I think it is fair to say that the van der Waals force is understood very well from the underlying QED.  Now we don't have that counterpart in QCD, but I'm reminded of some work by Nathan Isgur from the Jefferson Lab of some 25 years ago or so who made a brave attempt to describe the nucleon-nucleon force as a six body quark system. It was only partially successful but it seems to me with our present technology and improved computing abilities that it is worth revisiting that perhaps at the level of maybe meson-meson interactions which would be starting from a four quark system. That's all that I wish to say.

{\bf Bruno Juli\'a-D\'iaz} (Universidad de Barcelona):  I just wanted to make a comment because no one has mentioned here the recent calculations in lattice QCD which were mentioned by Frank Wilczek  in ``News and Views.''   This is a  calculation  by Ishii  and collaborators \cite{Is07} where they claim they are able to construct the nucleon-nucleon potential from lattice QCD. And Frank  Wilczek wrote explicitly ``I am very happy now that the theory I worked on is able to understand the nucleon-nucleon interaction.'' So this is actually connected to the last part of the first question:  are any of the methods ``transcendent.''  Don't you think that the one that is probably going to be ``transcendent'' is the nucleon-nucleon force which is obtained from lattice QCD and which, in some sense, are really connected to the equations of motion of QCD?

{\bf Gross}: I have to apologize because the organization of the conference is such that the lattice talk is on Thursday.  I just thought that it was too much to try to bring such a complex subject into this discussion, but of course I left out a very important subject. Nemura is here. We have to hear his talk and we have to have a discussion then about those issues.  Maybe he could make a comment now if he wants to. It certainly is important to fold that into our discussions. It really has been overlooked in our discussions so far. Do you want to make a comment now or would you rather wait. O.K., go ahead.

{\bf Hidekatsu Nemura} (Tohoku University):  An interesting point in the lattice QCD effort is to introduce strangeness in order to construct the hyperon-nucleon or hyperon-hyperon interaction. There is a large ambiguity. For example, how can we obtain the potential for the deuteron from  lattice QCD? This is a most important examination of lattice QCD. After understanding the nucleon-nucleon sector, we can  probably obtain a realistic hyperon-nucleon or hyperon-hyperon potential from lattice QCD.  So first we have to find techniques or a procedure for using lattice QCD to obtain the physical potential. 

{\bf Gross}: I think that is a very good point. We haven't talked about strangeness at all during this discussion and maybe we'll close this particular question with each of the panelists  making a very short comment about what they think about strangeness in the context of their own point of view. 

{\bf Cohen}: Strangeness can easily be included in large $N_c$ and  actually SU(2$N_f$) symmetries easily include strangeness.  I just want to make a comment about the lattice though.   In some sense one very much would like reliable lattice data for potentials. It is also worth observing there are some deep theoretical problems in trying to formulate this problem on the lattice. That is, the lattice actually computes correlation functions. It doesn't compute potentials directly and there is an additional theoretical step in trying to turn this into potentials. One reason you know that, has been mentioned here many times. Potentials aren't physical. You can change potentials and get the same observables by shifting things into different off-shell behaviors,  and so no unique potential exists as a matter of principle.  It's impossible to derive the unique potential from QCD on the lattice. So the real question is getting physical observables from the lattice which you can somehow relate to potentials.

{\bf Gross}: Good point. Do the other members of the panel want to say something about that.

{\bf Epelbaum}: Well, basically I agree with the statement, so I do believe that it is maybe more useful to have an interface based on calculating observables.

{\bf Gross}: Yes, but the coefficients of chiral perturbation theory ... yes, go ahead.

{\bf Machleidt}:   One more comment on the lattice.  Presently, two sets of lattice calculations exist in the world. One is the Japanese calculations which are very respectable calculations. The other set of calculations are called, I think, the nuclear physics lattice collaboration. Martin Savage and others are members and there they calculate observables, such as the scattering length, and then try to connect them to chiral perturbation theory results.   There you see directly a connection between lattice QCD and chiral perturbation theory and I think that is a very nice thing. 

{\bf Gross}: We have a few more comments.   Ru\' iz-Arriola first and then Sauer.

{\bf Ru\' iz-Arriola}: I completely disagree that a potential cannot be given a proper meaning.  Because we are in the infinite mass limit the potential is well defined. It's like saying the van der Waals force does not have any reality.   Because we don't know the short-range path we think that the long range cannot be determined.   I think that for very very heavy particles the potential is well defined.


{\bf Sauer}: Well, I'm coming back to the nuclear force.   It is important that we have a good connection between the two and three nucleon forces and QCD.   For me what is most important here is that you show us in chiral effective field theory potential forms, especially for the three nucleon force, which we would not have thought of before in a purely phenomenological approach. I think that is helpful for applications. 

I have a question for Evgeny which is really very naive.  You see, I admit I'm scared by all these diagrams. But you showed me something else. You showed us today this chiral window and that reminded me very much of the general approach that we always had for the two-nucleon force: first the long range pion exchange, then the intermediate range attraction which is mostly correlated two pion exchange, and then short range repulsion. You seem to indicate that behind all those horrible diagrams (which have a mathematical beauty) you are working very much in the region of intermediate range attraction, and despite all the contact terms you have, the short range, whatever it is,  seems to remain phenomenological. Is that somehow a correct view or do I misunderstand something?

{\bf Epelbaum}:   Yes, maybe I can  answer this question. I made the point that the long range contributions are perhaps most interesting in the sense that we can really learn something about them from QCD.  They ``feel'' chiral symmetry and we actually are able to calculate the corresponding potentials unambiguously. Now the short range contributions of course also do exist, but in chiral perturbation theory we just parameterize them in the most general way, so in this sense you don't get any gains out of chiral symmetry.

{\bf Gross}: Your 24 constants are the short range physics. Is that correct?

{\bf Epelbaum}: That's correct, yes. But I would also like to comment about your statement on the three-body interactions.  In fact, chiral effective theory can do more than just give you an idea about                  spin-isospin structure. Actually I have shown a picture where the profile function was indicated for a specific operator and these profile functions are unambiguous predictions of the theory without any adjustable parameters. So this long range behavior is, again, extremely strongly constrained due to the chiral symmetry of QCD and that is what we really can calculate.

{\bf Gross}: Another comment/question over here.

{\bf Alfred Stadler} (University of \'Evora):  I would like just to comment also on this question about potentials being somehow unphysical. I also don't quite agree with that view. I mean, first of all, potentials are hermitian operators so they must be observable. But I think there is a confusion of two different aspects here. You can measure potentials at large distances and, in fact, the first talk this morning showed how you can measure potentials and manipulate them at will, so there is nothing wrong with the potential by itself.  I think what everybody means is that at very short distances you just can't measure it directly. So, it is just this kind of ambiguity, you can't determine it from the data.  I think that lattice calculations might give you a handle on this part because effectively what one is doing there is simulating this measuring process. You have these configurations of two-three quark clusters at certain distances and you measure  the force between those two clusters and then you vary the distance.

{\bf Cohen}: Actually I disagree with that for the following reason -- when I say a potential I'm referring to an $NN$ potential, right, and the problem there is exactly what do you mean by $N$?  $N$ is not to be random quarks. $N$ is three quarks with all kinds of collations and exactly which operator you put into QCD will determine the exact thing you get out. That's what I meant by unphysical. The problem is not physical observables. The problem is defining what you mean by your fields and since the fields aren't fundamental in the theory you have to make some kind of ansatz. At long distances I should note it is completely unambiguous, and that's simply because no matter what you do you will be dominated by pion exchange.

{\bf Gross}: O.K. Next.

{\bf Stanislaw Kistryn} (Jagiellonian University, Cracow):  Perhaps I change the topic a bit by addressing the second question. I am an experimentalist. We measure very different samples of three and four nucleon systems and my dream is always that we do that with the purpose to advance theories, tests the models or theories, and provide finally the description which can also be used in other fields of nuclear physics. Well, what Peter Sauer says, let's study, for example,  collisions of two nuclei close to the coulomb barrier, low energy so every theory should be fine.  People use coupled channels, coupled reaction channels, continuum discretized coupled channels, to describe features of the reaction, to study the reaction mechanism. And I think the task or obligation of the few-body community would be to provide at least one piece of knowledge which is free of handles. That is, we say that the nucleon-nucleon or few-nucleon reaction is so, and one cannot turn any knobs to improve the agreement with the data. And that I would say is also important for other fields, such as spallation processes or multi-fragmentation processes. We should provide a final picture of the nucleon-nucleon interaction perhaps with three-nucleon forces, four-nucleon forces (which are small), but they are important to model the reaction mechanisms of more complicated systems.

{\bf Kievsky}: I would like to make a brief comment.  In my previous comment I suggested that by studying nuclear forces  we can do correctly  nuclear structure and nuclear physics.  But I want to stress what  Giuseppina  said before -- there is a very close relation between the study of nuclear forces and few-body physics and nuclear physics. The deeper our study of nuclear forces the deeper is the relation between the people studying  nuclear forces and those doing few body physics.   Of course, the study of the three-nucleon force in chiral perturbation theory is one of consistency.  But there is another argument which is that in few body physics we are not able to describe the asymmetries, for example,  proton-deuteron or proton-$^3$He.   I think this stimulates the study of three-nucleon forces a lot, because if we has a pretty good description the effort to go farther and farther  would not be so intense as is now.   So I think the collaboration now between the study of nuclear forces and few-body physics is more intense. This is my personal opinion.

{\bf Gross}: That actually leads to what I was going to do next. We are coming close to the end. I thought I would invite the panelists each to speak for about a minute or two and summarize what they think the discussion has been about. I was going to try to do that but I feel I'm too involved in some parts of it to do it. They are a little more neutral I think. So would the panelists be willing to do that? It might be interesting. 

{\bf  Epelbaum}:  I will just say a couple of words from my personal perspective. I think that perhaps the most interesting field in the near future will be three-body interactions, or three-nucleon interactions, just because there are several puzzles and it is really time to try to address those issues. So we are working on these interactions and on implementing these interactions.  The dream or the hope is that once we have succeed we'll produce, on the computer or on a disk, matrix elements in the partial wave decomposition.  Of course, once we are done with this we would be happy to share those ingredients with any interested group.

{\bf Cohen}:  I just want to take a very broad view on this whole question. I think it was clear that there were two different sets of interests. I think Peter Sauer sort of accurately gave the ``who cares'' view which is actually a very respectable one.   Basically, it is very, very, hard to connect QCD directly to the properties of nuclei. The key point is that one is hoping here to learn at least qualitatively how to connect things in nuclear physics to QCD. From a practical point of view, large $N_c$ is not terribly useful except for that purpose. Chiral perturbation theory has a lot of promise in the sense that apart from being connected to QCD, it gives an organizing principle by which one has some hope of learning a priori what's going to be large and what's going to be small, particularly with regard to external currents.  And that can be very, very, useful in various processes where you shoot in a photon and knock out a pion, and the like. So there is some sort of practical benefit but I think the bigger question of how nuclear phenomenon is connected to QCD is still fairly uncertain and ultimately, except through lattice calculations, it's going to remain that way. 
      
{\bf Gross}: Want the last word?
      
{\bf Machleidt}: To put this also in a relatively broad framework: we are physicists and not nuclear engineers, so we do basic science. The first goal is  to look for the fundamentally correct theories, and that has both an intrinsic value and an extrinsic value.  The intrinsic value is that, when we find the right theory, that is a value by itself.  Concerning the nuclear force, I think that our goal is to find the one that is on the best fundamental grounds, and that is presently the chiral perturbation approach, because it  has a much more obvious connection to QCD than all the others, even though respected, approaches. 

Then, when it comes to the extrinsic aspect of the theories, that means the applications, then at first glance the traditional meson theory potentials and the chiral potentials do equally well. They do not give exactly the same results, but the differences are actually only off-shell differences and you know that off-shell differences are not an aspect. However, when it comes finally to more sophisticated applications, nuclear structure calculations where three-body forces, currents and other aspects play a role, then again, even in the extrinsic (which is the practical) aspect, the chiral perturbation theory has an advantage because it also provides these other technical aspects like three body forces, etc., in a more systematic and reliable way. 
      
{\bf Franz Gross}:   Well I think we've used up our time. I appreciate all the comments. We'll do our best to get them into the proceedings in the correct way, and I think we should all applaud ourselves.  


\begin{acknowledgements}

Tom Cohen gratefully acknowledges support of the U.S. Department of Energy under grant number DE-FG02-93ER-40762.  Evgeny Epelbaum gratefully achnowledges support of the Helmholtz Association under grant numbers VH-NG-222 and
VH-VI-231.  Franz Gross was supported by Jefferson Science Associates, LLC under U.S. DOE Contract No.~DE-AC05-06OR23177.  R. Machleidt was supported in part by the Department of Energy under Grant No. DE-FG02-03ER41270.
\end{acknowledgements}


\end{document}